\documentclass[
reprint,
amsmath,
amssymb,
aps,
superscriptaddress,
nofootinbib,
preprintnumbers,
prd
]{revtex4-2}
\usepackage{hyperref}
\hypersetup{colorlinks=true, citecolor=blue, urlcolor=blue, linkcolor=blue}
\usepackage[T1]{fontenc}
\DeclareRobustCommand{\VAN}[3]{#2}
\let\VANthebibliography\thebibliography
\def\thebibliography{\DeclareRobustCommand{\VAN}[3]{##3}\VANthebibliography}

\def\thebibliography{\DeclareRobustCommand{\VAN}[3]{##3}\VANthebibliography}
\usepackage{graphicx}	
\usepackage{amsmath}	

\usepackage{amssymb}	
\usepackage{bm}		

\usepackage{ae,aecompl}
\usepackage{newtxtext,newtxmath}
\usepackage{xurl}

\usepackage{dcolumn}    		
\usepackage{aas_macros}
\usepackage{tipa} 
\usepackage[table]{xcolor}

\newcommand{\kmax}{\ensuremath{k_{\rm UV}}}
\newcommand{\kmaxz}{\ensuremath{k_{\rm UV}(z)}}
\newcommand{\kmaxzero}{\ensuremath{k_{\rm UV,0}}}
\newcommand{\lmax}{\ensuremath{\ell_{\rm max}}}
\newcommand{\Lmax}{\ensuremath{L_{\rm max}}}
\newcommand{\hmpc}{\ensuremath{\,h\,{\rm Mpc}^{-1}}}

\renewcommand{\selectlanguage}[1]{} 

\begin{document}

\title[EFT3]{Toward one- and two-loop EFTofLSS for Simons Observatory CMB lensing}

\author{Evan Saraivanov}
\email[First author: ]{evan.saraivanov@stonybrook.edu}
\affiliation{Department of Physics and Astronomy, Stony Brook University, Stony Brook, NY 11794, USA }
\affiliation{C.N. Yang Institute for Theoretical Physics, Stony Brook University, Stony Brook, NY 11794, USA }
\author{Henrique Rubira}
\email[Second author: ]{henrique.rubira@lmu.de}
\affiliation{University Observatory, Faculty of Physics, Ludwig-Maximilians-Universit\"at, Scheinerstr. 1, D-81679 München, Germany}
\affiliation{Kavli Institute for Cosmology Cambridge, Madingley Road, Cambridge CB3 0HA, UK}
\affiliation{Centre for Theoretical Cosmology, Department of Applied Mathematics and Theoretical Physics
	University of Cambridge, Wilberforce Road, Cambridge, CB3 0WA, UK}
\author{Vivian Miranda}
\affiliation{C.N. Yang Institute for Theoretical Physics, Stony Brook University, Stony Brook, NY 11794, USA }

\begin{abstract}
{
Cosmic microwave background (CMB) lensing observations are becoming sensitive to non-linear structure formation and baryonic feedback scenarios. The effective field theory of large-scale structure (EFTofLSS) provides a systematic description of matter clustering in the mildly non-linear regime, accounting for baryonic effects through counterterms and offering a consistent framework for cross-correlations with other observables. We perform a Fisher forecast for the joint Simons Observatory temperature, polarization, and CMB lensing likelihood using noiseless data vectors with different baryonic feedback scenarios, modelling non-linear structure formation using the one- and two-loop EFTofLSS. We find that, with a single additional parameter, obtained through a principal-component analysis of the EFTofLSS counterterms, both the one- and two-loop EFTofLSS provide a good fit across the baryonic feedback scenarios considered. This parameter is closely aligned with the effective AGN heating temperature $T_{\rm AGN}$. Both loop orders reduce cosmological parameter biases, with only a small loss in marginalized precision at fixed scale cuts. We find that the one-loop EFTofLSS yields unbiased estimates over a range of scales similar to the dark-matter-only model, while the two-loop EFTofLSS extends this range to smaller scales and accommodates different feedback scenarios more robustly.  When each model is evaluated at the smallest scales for which its cosmological parameter estimates remain unbiased, the two-loop EFTofLSS yields stronger constraints than the dark-matter-only model.
}
\end{abstract}

\maketitle

\section{Introduction}

The Simons Observatory (SO)~\cite{SimonsObservatory:2025wwn} and other CMB surveys such as the South Pole Telescope~\cite{SPT-3G:2024qkd,SPT-3G:2014dbx} and the Atacama Cosmology Telescope~\cite{AtacamaCosmologyTelescope:2025blo,AtacamaCosmologyTelescope:2025nti,LaPosta:2026tzv} measure the cosmic microwave background (CMB) at high multipoles with unprecedented precision. The CMB signal is distorted by gravitational lensing of CMB photons by the large-scale structure (LSS) of the Universe~\cite{LEWIS_2006}. Gravitational lensing affects a wide range of multipoles by smoothing the acoustic peaks and transferring power into the damping tail, where the lensing contribution can dominate over the unlensed CMB signal.
CMB lensing is an important probe of late-time structure formation and can constrain dark matter and dark energy physics through their effects on structure growth and the expansion history~\cite{ACT:2025qjh}.

It has been shown, however, that for SO and next-generation surveys, inconsistent modelling of non-linear clustering and baryonic effects can systematically bias cosmological parameter inference 
~\cite{McCarthy_2022,trendafilova2025endeasyphenomenologycmb,McCarthy:2020dgq}. The effective field theory of large-scale structure (EFTofLSS)~\cite{Baumann_2012,Carrasco_2012,Carrasco_2014,Konstandin_2019,ivanov2022effectivefieldtheorylarge} can mitigate this by providing a consistent framework for describing the matter power spectrum in the mildly non-linear regime. It combines perturbative calculations of long-wavelength modes with counterterms that parameterize the large-scale effects of unresolved small-scale dynamics, including both non-linear gravitational evolution and baryonic physics~\cite{Braganca:2020nhv}. While marginalizing over the counterterm coefficients can broaden cosmological parameter constraints, higher-order computations can extend the range of scales accurately described by the theory, providing access to additional cosmological information. 

Recent work has explored two-loop predictions for two-point statistics in galaxy clustering and weak lensing~\cite{Saraivanov:2026sxc,Chen:2026usz,Ivanov:2026zos,philcox2026redshiftspacegalaxypowerspectrum,Bakx:2025cvu,bakx2026galaxybiasrenormalizationtwoloop,Bakx:2025jwa}, making two-loop calculations a promising route to extracting more cosmological information from smaller scales within the perturbative regime.
In this manuscript, we apply the EFTofLSS to a joint forecast for SO lensed CMB temperature and polarization power spectra and the CMB lensing power spectrum.
We investigate whether one-loop and two-loop EFTofLSS predictions for the non-linear matter power spectrum yield unbiased cosmological parameter estimates and assess the precision of these constraints under different scale cuts.

We parameterize the redshift dependence of the counterterms using a spline basis and further reduce the dimensionality of the parameter space by applying principal component analysis (PCA) to the spline coefficients. Within this counterterm parameterization, only one additional nuisance parameter is needed at either loop order to capture the dominant effect of the counterterms on CMB lensing. We find this parameter to be closely aligned with the active galactic nuclei heating temperature, which is one source of baryonic feedback.  

The paper is organized as follows. In Section~\ref{sec:eftoflss}, we review the EFTofLSS up to two-loop order and outline the counterterm parameterization used in our forecasts.
In Section~\ref{sec:cmb_lensing}, we review the lensing power spectrum and the leading-order lensing contributions to the CMB temperature and polarization spectra. In Section~\ref{sec:data_analysis}, we explain the priors and how we determine the scale cuts used in the analysis. In Section~\ref{sec:forecast}, we present forecasts, comparing the cosmological constraints and parameter biases obtained using dark-matter-only spectra and one-loop and two-loop EFTofLSS predictions.

\section{The EFT of large-scale structure}
\label{sec:eftoflss}

The effective field theory of large-scale structure~\cite{Baumann_2012,Carrasco_2012,Carrasco_2014,Konstandin_2019,ivanov2022effectivefieldtheorylarge} (EFTofLSS) is an analytic description of cold dark matter in the mildly non-linear regime where density perturbations remain small. This comes at the cost of additional nuisance parameters ({\it counterterms}) that parameterize unknown small-scale physics that was integrated out of the theory. The corrections are suppressed by the characteristic length scale of large-scale structure, the non-linear scale $k^{-1}_{\rm NL}$. 

For the one- and two-loop matter power spectra in the EFTofLSS, we have the following expressions:
\begin{equation}
\begin{split}
    P_{\rm EFT,1L}(k,z) &= P_{\rm L}(k,z) + P_{1}(k,z) - P_{1,{\rm c.t}}(k,z) \,,\\
    P_{\rm EFT,2L}(k,z) &= P_{\rm L}(k,z) + P_{1}(k,z) + P_{2}(k,z) \\
    &\qquad - P_{2,\rm{c.t}}(k,z)\,,
\end{split}
\end{equation}
where $P_{\rm L}$ is the linear power spectrum and $P_{1}$ and $P_{2}$ are the standard perturbative loop corrections to the power spectrum \cite{Bernardeau:2001qr}. The one-loop counterterm $P_{1,{\rm c.t}}$ is given by the effective sound speed contribution
\begin{equation} \label{eq:1lct}
    P_{1,{\rm c.t}}=2 c_{s,(1)}^2 (z) \left(\frac{k}{k_{\rm NL}}\right)^2 P_{\rm L}(k,z)\,,
\end{equation}
with $c_{s,(1)}^2 (z)$ a free coefficient that depends on redshift.
The two-loop counterterms we consider for the power spectrum are given by 
\begin{equation} \label{eq:2loopct}
\begin{split}
    P_{2,{\rm c.t}} & = 2(c_{s,(1)}^2(z)+c_{s,(2)}^2(z)) (k/k_{\rm NL})^2 P_{\rm L}(k,z) \\
    &\quad+ c_{s,(1)}^2(z) (k/k_{\rm NL})^2 P_{1}(k,z) \\
    &\quad+ (c_{s,(1)}^4(z) - 2c_4(z)) (k/k_{\rm NL})^4 P_{\rm L}(k,z) \\
    &\quad+ 2c_{\rm quad}(z) (k/k_{\rm NL})^2 P_{\rm quad}(k,z) \\
    &\quad- c_{J}(z) (k/k_{\rm NL})^4\,,
\end{split}
\end{equation}
with $P_{\rm quad}(k) = \int d^3\mathbf{q}/(2\pi)^{3} P_{\rm L}(q) P_{\rm L}(|\mathbf{k}-\mathbf{q}|) F_2(\mathbf{q}, \mathbf{k}-\mathbf{q})$ being the only term quadratic in the density field we consider. For recent discussions on the expanded set of two-loop matter counterterms, see \cite{Anastasiou:2025jsy, bakx2026galaxybiasrenormalizationtwoloop}, with the other counterterms being suppressed relative to the ones in Eq.~\ref{eq:2loopct} \cite{Anastasiou:2025jsy}.
We also include the stochastic contribution $c_{J} (k/k_{\rm NL})^4$. 
We use neural networks to emulate each of the terms needed for the IR-resummed power spectrum, the details of which are given in App.~\ref{sec:emulator}.

\begin{figure}
    \centering
    \includegraphics[width=\columnwidth]{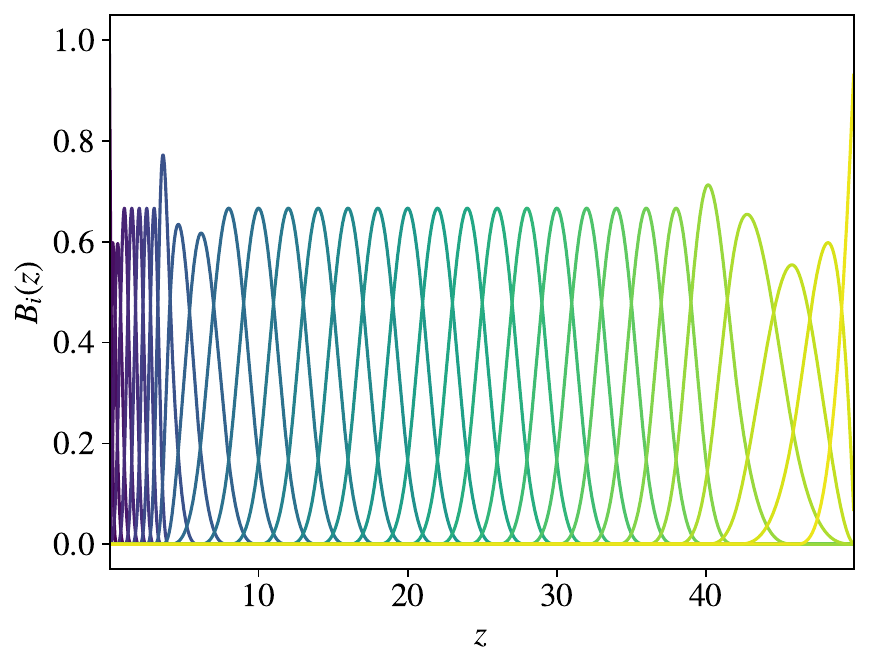}
    \caption{
    The cubic spline basis functions adopted for the counterterms. The computation of the full Wilson coefficient is a linear operation on amplitude coefficients as outlined in Eq.~\ref{eq:basis}. 
    }
    \label{fig:spline_basis}
\end{figure}

The Wilson coefficients above, $c_{s,(1)}^2$, $c_{s,(2)}^2$, $c_4$, $c_{\rm quad}$, and $c_{J}$, acquire a time dependence due to the EFTofLSS's non-locality in time that is separate from the time evolution of $k_{\rm NL}$. For a power-law universe with $P(k) \propto [D(z)]^2(k/k_{\rm NL})^n$, with $D(z)$ being the linear growth factor, the scaling $k_{\rm NL}\propto [D(z)]^{-2/(3+n)}$ captures the bulk of the time evolution. While this scaling provides a very good approximation, we take a more model-agnostic approach, modelling the full time evolution of the Wilson coefficients, which, for $P_{2,{\rm c.t}}$ we reparameterize as
\begin{equation}\label{eq:counterterms}
\begin{split}
    \bar{c}_{s}^2      &= \left(c_{s,(1)}^2 + c_{s,(2)}^2\right)/k_{\rm NL}^2\,,\\
    \bar{c}_{s,(1)}^2  &= c_{s,(1)}^2/k_{\rm NL}^2\,,\\
    \bar{c}_4          &= \left(c_{s,(1)}^4/2 - c_4\right)/k_{\rm NL}^4\,,\\
    \bar{c}_{\rm quad} &= c_{\rm quad} / k_{\rm NL}^2\,,\\
    \bar{c}_{J}        &= 10^{-3}c_{J} / \left(k_{\rm NL} [D(z)]^{3/2}\right)^4\,.
\end{split}
\end{equation}
Notably, we factor out a growth factor dependence from the stochastic term so that it scales similarly to other two-loop terms.\footnote{For our application, this also serves a numerical purpose. The redshift modelling of the power spectrum must be extended up to $z=50$ for CMB lensing, so if not for the growth factor in the stochastic term, the data would mostly constrain the high-redshift behaviour of $\bar{c}_J$ rather than the low-redshift behaviour of all of the terms. At high redshifts, where the other terms drop away, requiring a non-negative power spectrum puts strict lower bounds on the value of $c_{J}$, making it difficult to compute a linearized forecast.}  

We model the redshift dependence of the Wilson coefficients using a cubic spline basis, shown in Fig.~\ref{fig:spline_basis}. The basis is a cubic $B$-spline with a knot vector chosen so that $z<4$, which contains the peak of the CMB lensing kernel, has a basis function peak every $\Delta z\approx 0.5$ and $z\geq4$ has a peak every $\Delta z\approx2$. This spacing gives us $32$ basis functions, and therefore $32$ free parameters per Wilson coefficient. These functions generate a partition of unity on the interval $z\in[0,50]$.
Modelling up to $z=50$ ensures that the EFTofLSS covers the range where the CMB lensing window function has meaningful support, and that the matter power spectrum is sufficiently linear that perturbative modelling is not necessary at higher redshift. The full Wilson coefficients are then determined by
\begin{equation}\label{eq:basis}
    c_{X}(z) = \boldsymbol{c_X}\cdot\boldsymbol{B}(z) = \sum_{i=1}^{32} {c}_{X,i} B_i(z)\,,
\end{equation}
with $c_{X}$ labelling one of the coefficients in Eq.~\ref{eq:counterterms}. Thus, the free parameters of our two-loop model are the 160 ${c}_{X,i}$. The data is of course not sensitive to all of these parameters, so we reduce the dimensionality using a principal component analysis (PCA), as outlined in Sec.~\ref{sec:parameters}. Note that, in contrast with~\cite{Saraivanov:2026sxc}, which fitted a component of the counterterms in dark-matter-only emulators, here we let the counterterms be completely determined by the free parameters. This is done to minimize the reference to external non-linear models up to 
\begin{equation}\label{eq:kmaxscaling}
    \kmax (z)= {\kmax}_{,0} /D(z)\,,
\end{equation}
the highest wavenumber we consider in the analysis and for the fit of the counterterms. 

An additional caveat for lensing observables is the divergent nature of some of the EFTofLSS counterterms in Eq.~\ref{eq:2loopct} for $k>\kmax$. At two loops, the stochastic term, the four-derivative term, and the sound-speed-sound-speed autocorrelation term are particularly problematic at large wavenumbers. On the other hand, if we were to truncate the power spectrum at $k=\kmax$ (i.e., set $P(k>\kmax)=0$), we would systematically underpredict the projected lensing signal due to the neglected power at small scales. Thus, we must make some choice of extrapolation until the power spectrum is sufficiently suppressed that smaller scales can be neglected. We extrapolate the power spectrum using a reference non-linear model according to
\begin{align}\label{eq:extrap}
    R(k \leq \kmaxz,z) &= R_{\rm EFT}(k \leq \kmaxz,z) \,, \\
    R(k > \kmaxz,z) &=  \frac{R_{\rm EFT}(\kmaxz,z)}{R_{\rm ref}(\kmaxz,z)} R_{\rm ref}(k > \kmaxz,z)\,, \nonumber
\end{align}
with $R$ being the ratio of the non-linear spectrum to the linear spectrum. In our forecast, we use the $\Lambda$CDM non-linear ratio to extrapolate the power spectrum for $k>\kmax$. This is slightly different from~\cite{Braganca:2020nhv,Saraivanov:2026sxc}, since the halo model we use for the extrapolation assumes $\Lambda$CDM rather than $w_0w_a$CDM. However, in the spirit of the EFT, our final results should be insensitive to these UV $k$-modes. We address this problem in Sec.~\ref{sec:scale_cuts}, where we interpret different choices of reference models as a theoretical uncertainty to be marginalized over using scale cuts.

\section{CMB Lensing}
\label{sec:cmb_lensing}

The CMB photons are deflected as they traverse the large-scale structure~\cite{LEWIS_2006}.
These deflections are determined by the dimensionless power spectrum of the Weyl potential $\Psi_+ = (\Psi + \Phi)/2$, where $\Psi$ and $\Phi$ are the scalar perturbations of the metric in the Newtonian gauge. This spectrum is related to the matter power spectrum by
\begin{equation}
    P^{\Psi_+}(k,\chi) =  \frac{9}{8\pi^2k}\Omega_{\rm m}^2(\chi) a^4(\chi) H^4(\chi) P(k,\chi)\,,
\end{equation}
where $H$ is the physical Hubble rate, $a$ is the scale factor, and $\Omega_{\rm m}$ is the matter density fraction.

Within the Born approximation, the lensing potential $\phi$ is defined by
\begin{equation}
    \phi(\hat{\mathbf n}) = 2\int\limits_0^{\chi_*}d\chi\,W(\chi) \Psi_+(\chi\hat{\mathbf n},\eta_0-\chi)\,, 
\end{equation}
where $\eta_0$ is the present conformal time, $W(\chi)= (\chi - \chi_*)/(\chi_*\chi)$ is the CMB lensing window function, and $\hat{\boldsymbol{n}}$ is the unit vector along the line of sight. Projecting the Weyl power spectrum along the line of sight, the angular power spectrum of $\phi$ is given by
\begin{equation}
\begin{split}
    C_L^\phi =& 16\pi \int\limits_{-\infty}^{\infty} d \log k \int\limits_0^{\chi_*} d\chi' \int\limits_0^{\chi_*} d\chi \,\, W(\chi) W(\chi') \\ 
    &\quad \times j_{L}(k\chi) j_{L}(k\chi') P^{\Psi_+}(k, \chi, \chi')\,,
\end{split}
\end{equation}
with $\chi_*$ the comoving radial distance to the last-scattering surface.
We use $P^{\Psi_+}(k,\chi,\chi') \simeq \sqrt{P^{\Psi_+}(k,\chi)P^{\Psi_+}(k,\chi')}$ for the unequal-time Weyl power spectrum.

The resulting correction to the CMB power spectrum is obtained via the four-point correlation function involving two temperature/polarization fields and two lensing potential fields. Assuming the CMB and the lensing potential are uncorrelated, the lensed temperature autocorrelation spectrum is given by
\begin{equation}
\label{eq:delta_tt}
\begin{split}
    \Delta C_{\ell}^{TT} &= \int \frac{d^2 \ell'}{(2\pi)^2} \, [\ell' \cdot (\ell-\ell')]^2 \, {C_{\ell'}^{TT}}_{\rm unl.} C_{|\ell-\ell'|}^{\phi} \\
    &\quad- {C_{\ell}^{TT}}_{\rm unl.} \int \frac{d^2 \ell'}{(2\pi)^2} \, [\ell' \cdot \ell]^2 \, C_{\ell'}^{\phi}
\end{split}
\end{equation}
to leading order in $C_{L}^{\phi}$ and in the flat-sky approximation. Here, we use the notation where $\Delta C_\ell^{XY}\equiv C_{\ell,\rm lensed}^{XY}-C_{\ell,\rm unl.}^{XY}$ is the difference between the lensed and unlensed CMB spectra.

For the polarization spectra and cross spectra, there is an extra $\cos[2(\phi_{\ell'} - \phi_{\ell})]$ factor for each polarization field, so the flat-sky leading-order expressions are given by
\begin{equation}
\label{eq:delta_te}
\begin{split}
    \Delta C_{\ell}^{TE} &=
    \int \frac{d^2 \ell'}{(2\pi)^2} \,
    [\ell' \cdot (\ell-\ell')]^2 \,
    {C_{\ell'}^{TE}}_{\rm unl.} C_{|\ell-\ell'|}^{\phi}\\
    &\times \cos[2(\phi_{\ell'} - \phi_{\ell})]
    - {C_{\ell}^{TE}}_{\rm unl.}
    \int \frac{d^2 \ell'}{(2\pi)^2} \,
    [\ell' \cdot \ell]^2 \, C_{\ell'}^{\phi}
\end{split}
\end{equation}
and
\begin{equation}
\label{eq:delta_ee}
\begin{split}
    \Delta C_{\ell}^{EE} &=
    \int \frac{d^2 \ell'}{(2\pi)^2} \,
    [\ell' \cdot (\ell-\ell')]^2 \,
    {C_{\ell'}^{EE}}_{\rm unl.} C_{|\ell-\ell'|}^{\phi} \\
    & \times \cos^2[2(\phi_{\ell'} - \phi_{\ell})]
    - {C_{\ell}^{EE}}_{\rm unl.}
    \int \frac{d^2 \ell'}{(2\pi)^2} \,
    [\ell' \cdot \ell]^2 \, C_{\ell'}^{\phi}\,,
\end{split}
\end{equation}
with $\phi_{\ell}$ (not to be confused with the $\phi$ in the lensing spectrum) the azimuthal angle of the wavevector $\ell$. These polarization expressions assume negligible unlensed $B$-modes.

In Sections~\ref{sec:data_analysis} and~\ref{sec:forecast}, the data vector is constructed out of the binned $C_{\ell,{\rm lensed}}^{XY}$ spectra, as well as the binned lensing spectrum $C_{L}^{\phi}$. We compute the spectra using CAMB~\cite{Lewis:2026mif} with the accuracy settings outlined in App.~\ref{app:CAMBsettings}. These settings were used in~\cite{zhu2025attentionbasedneuralnetworkemulators} and produce negligible $\Delta\chi^2$ at cosmic variance precision. The linear matter power spectrum is computed using CAMB and the non-linear matter power spectrum ratio is provided by the EFTofLSS emulator and passed through CAMB's external non-linear ratio model. We use COBAYA~\cite{torrado_cobaya_2021} to connect the CAMB results to the SO likelihoods.

\section{Data and analysis}
\label{sec:data_analysis}

In this section, we describe the data vectors used in our CMB lensing forecasts. We first specify the cosmological parameters and their priors, then describe the principal-component parameterization of the EFTofLSS counterterms. Finally, we derive scale cuts for the lensed CMB temperature and polarization spectra and the CMB lensing spectrum as a function of $\kmaxzero$.

\subsection{Parameters and priors}\label{sec:parameters}

\begin{table}
    \centering
    \renewcommand{\arraystretch}{1.3}
    \begin{tabular}{|c|c|c|}
        \hline
        Parameter            & Value    & $\sigma_{\rm ref}$    \\\hline
        $\log(10^{10}A_s)$   & $3.04$   & $1.0\times10^{-2}$    \\
        $n_s$                & $0.966$  & $2.49\times10^{-3}$   \\
        $H_0$                & $66.7$   & $5.72\times10^{-1}$   \\
        $\Omega_{\rm b} h^2$ & $0.0224$ & $5.85\times10^{-5}$   \\
        $\Omega_{\rm c} h^2$ & $0.121$  & $6.42\times10^{-4}$   \\
        $\tau$               & $0.0543$ & $5.70\times10^{-3}$   \\
        $w_0$                & $-0.752$ & $5.55\times10^{-2}$   \\
        $w_a$                & $-0.729$ & $2.00\times10^{-1}$   \\
        \hline
    \end{tabular}
    \caption{Fiducial cosmological parameters adopted in this work. The background evolution parameters $H_0$, $\Omega_{\rm b}h^2$, $\Omega_{\rm c}h^2$, $w_0$, and $w_a$ are set to the DESI+DES-Y5 mean values, which we obtained by running an MCMC with a BBN-derived prior on $\Omega_{\rm b}h^2$~\cite{DESI:2025zgx}. The remaining parameters, $\tau$, $\log(10^{10}A_s)$, and $n_s$, are set to the Planck 2018 best-fit values~\cite{Planck:2018vyg}. The reference error bars, $\sigma_{\rm ref}$, are obtained from the dark-matter-only HMCode analysis with the $\kmaxzero=0.8\hmpc$ scale cuts.}
    \label{tab:fiducial_cosmology}
\end{table}

To demonstrate the EFTofLSS modelling of non-linearities for CMB lensing in Sec.~\ref{sec:forecast}, we perform a forecast for the Simons Observatory multifrequency TT, TE, and EE likelihood\footnote{\url{https://github.com/simonsobs/LAT_MFLike}} and the CMB lensing likelihood\footnote{\url{https://github.com/simonsobs/SOLikeT}}, using the covariance matrices provided therein. 

We generate noiseless data vectors for the lensed CMB TT, TE, and EE spectra and the $\phi$ lensing spectrum using the cosmological parameters in Tab.~\ref{tab:fiducial_cosmology}. The background evolution parameters $H_0$, $\Omega_{\rm b}h^2$, $\Omega_{\rm c}h^2$, $w_0$, and $w_a$ are derived from the mean of a joint DESI-DR2 BAO~\cite{DESI:2025zgx} and DES-Y5 supernova~\cite{DES:2024jxu} posterior with a BBN prior on $\Omega_{\rm b}h^2$~\cite{DESI:2025zgx}. $\log(10^{10}A_s)$ and $n_s$ are given by the Planck best-fit values~\cite{Planck:2018vyg}.
We consider non-linear matter power spectra computed with HMCode~\cite{Mead:2020vgs}, both without baryonic feedback and with the feedback prescriptions described below. For the data vectors with baryonic feedback, we apply the multiplicative correction to the dark-matter power spectrum from HMCode, which is then used for the non-linear lensing. For the feedback models, we use the correction from simulations up to their respective maximum redshift provided. For higher redshifts, we use the dark-matter-only spectrum from HMCode.

We quantify the parameter shifts induced by different non-linear prescriptions relative to the reference uncertainties $\sigma_{\rm ref}$ listed in the last column of Tab.~\ref{tab:fiducial_cosmology}, obtained from a dark-matter-only HMCode forecast using the scale cuts corresponding to $\kmaxzero=0.8\hmpc$ (see Sec.~\ref{sec:scale_cuts}) and the cosmological priors specified below. 
For $H_0$, $w_0$, $w_a$, $\Omega_{\rm b}h^2$, and $\Omega_{\rm c}h^2$, we adopt a Gaussian prior derived from DES-Y5 supernova~\cite{DES:2024jxu} and DESI BAO distance measurements~\cite{DESI:2025zgx}, centered at the fiducial values in Tab.~\ref{tab:fiducial_cosmology}. We determine the mean and covariance of this prior by running a Markov chain Monte Carlo analysis on the joint DES-Y5 and DESI BAO likelihood. This prior particularly helps constrain the Hubble constant and the dark energy equation-of-state parameters. We also include a Planck-derived Gaussian prior on the optical depth $\tau$, centered at $\tau=0.0543$ with standard deviation $\sigma^{\rm prior}_{\tau}=7.5\times10^{-3}$~\cite{Planck:2018vyg}, and we adopt uniform priors on $\log(10^{10}A_s)$ and $n_s$ over the ranges $[1.61,3.91]$ and $[0.8,1.2]$, respectively.

For the EFTofLSS parameters, we perform a PCA to reduce the dimensionality of the parameter space, similar to what was considered in \cite{Saraivanov:2026sxc}. For each of the $\kmaxzero$ in the range $[0.05,0.8]\hmpc$, we compute the Fisher matrix of the spline coefficients while keeping the cosmology fixed to Tab.~\ref{tab:fiducial_cosmology} and diagonalize the sum of the Fisher information of all of the $\kmaxzero$. We explicitly checked that choosing a $\kmaxzero$-dependent basis has no appreciable quantitative or qualitative effects on the results in Sec.~\ref{sec:forecast}. The PCA defines a new basis of functions, which are different for the one-loop and the two-loop EFTofLSS because of the different counterterm structure, and can be ranked by how constrained they are by the data. We then choose a number of PC parameters to keep and fix the remaining ones to $0$. 

This approach has a few advantages and disadvantages compared to the method in~\cite{Saraivanov:2026sxc}, which considered a different PCA basis for different $\kmax$. The main advantage is that keeping a consistent basis across $\kmax$ allows us to observe overfitting more directly than before, as we comment in Sec.~\ref{sec:overfitting}. However, given that the PCA basis is no longer the optimal basis for a particular $\kmax$, we may be losing some constraining power after marginalizing over the un-fixed PCs. As such, the constraining power we show for a given number of principal components is potentially understated compared to a more optimal basis derived from the Fisher matrix at a particular $\kmax$. We have verified that this effect is small enough that it does not affect the results.

Following the simulation-based approach of~\cite{Saraivanov:2026sxc}, we calibrate the EFTofLSS priors by fitting the counterterms to matter power spectra separately using different dark energy parameters and baryonic feedback prescriptions, with the other cosmological parameters fixed at their fiducial values.
We vary $w_0$ and $w_a$ over the ranges $[-1.3,-0.7]$ and $[-0.3,0.3]$, respectively, and consider baryonic feedback from HMCode with $T \equiv \log_{10}[T_{\rm AGN}/(1\,\mathrm{K})]=7.8$, as well as feedback models from
BAHAMAS~\cite{McCarthy_2018},
OWLS-AGN~\cite{Le_Brun_2014},
TNG~\cite{Pillepich_2017,Springel_2017,Naiman_2018,Marinacci_2018,Nelson_2017},
EAGLE~\cite{Crain_2015,Schaye_2014},
Antilles~\cite{Salcido_2023},
MB2~\cite{Khandai_2015},
and Illustris~\cite{Genel_2014,Vogelsberger_2014}.
For the hydrodynamical prescriptions, we construct the target spectrum as $P_{\rm target}(k,z)=P_{\rm HMCode,DMO}(k,z)R_{\rm bar}(k,z)$, where $P_{\rm HMCode,DMO}$ is the dark-matter-only HMCode prediction at the chosen cosmology and $R_{\rm bar}$ is the ratio of the hydrodynamical to the corresponding dark-matter-only simulation spectrum. We assume $R_{\rm bar}$ is cosmology independent.
For each target spectrum, we fit the EFTofLSS counterterms using the fractional residual $[P_{\rm EFT}(k,z)-P_{\rm target}(k,z)]/P_{\rm target}(k,z)$.
The fits extend to $\kmax(z)=0.3\,D^{-1}(z)\hmpc$~\cite{Foreman:2015lca}, where $D(z)$ is the linear growth factor.
We then project the fitted counterterm functions onto the PCA basis and compute the covariance of the resulting amplitudes across the calibration spectra. We make the prior more conservative by neglecting correlations and inflating the variances by a factor of $25$, ensuring that all of the models are within about $2\sigma$ of the prior mean of $0$.

At $\kmaxzero \lesssim 0.3 \hmpc$, this prior becomes highly informative on the PC amplitudes. Since the prior for the EFTofLSS is uncorrelated with cosmology, the correlation between the cosmology and PC amplitudes becomes weak, and the apparent marginalized constraining power on the cosmological parameters increases as data is removed. Hence we add a $\kmaxzero$-dependent variance to the prior that corresponds to the fitting error, ensuring that the prior is never more informative than fitting the power spectrum, or alternatively that the prior at $\kmaxzero<0.3\hmpc$ is not entirely set by prior information calibrated at $\kmaxzero=0.3\hmpc$.

We keep the CMB foreground parameters fixed in our forecast (which is also the case for the reference uncertainties in Tab.~\ref{tab:fiducial_cosmology}), allowing us to isolate the impact of the EFTofLSS on the fit quality and its ability to mitigate cosmological parameter biases. The statistical significance of the parameter biases may therefore be overstated relative to an analysis that marginalizes over foreground uncertainties.

\subsection{Scale cuts}
\label{sec:scale_cuts}

\begin{figure}
    \centering
    \includegraphics[width=\columnwidth]{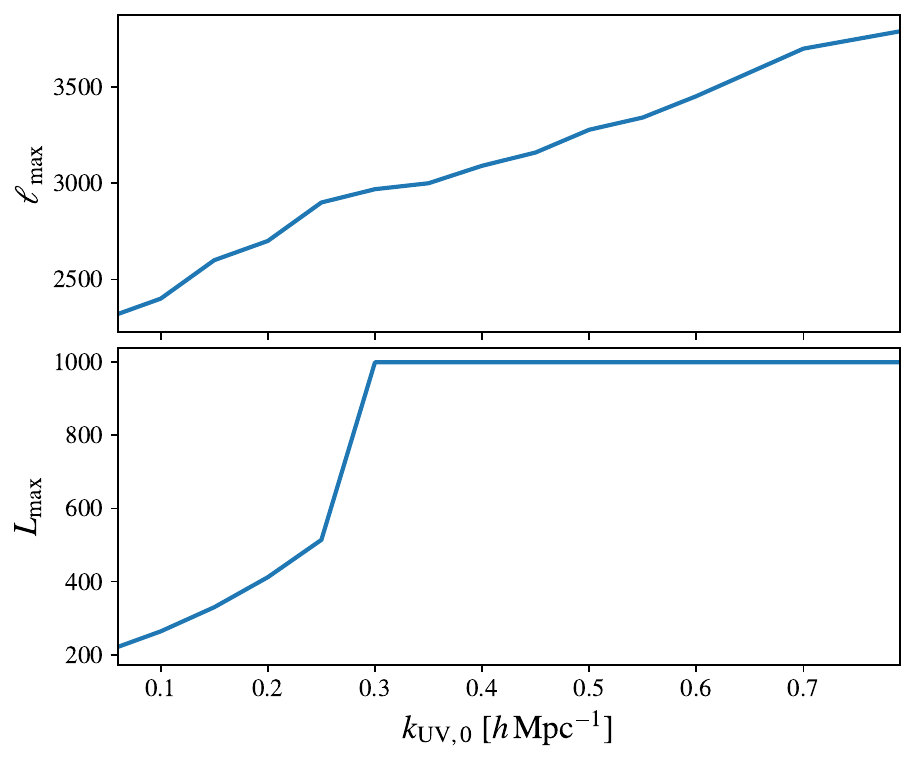}
    \caption{
    The scale cuts adopted in this work as a function of $\kmaxzero$ using the method outlined in Sec.~\ref{sec:scale_cuts}. The upper panel is the maximum multipole for the TT, TE, and EE spectra, and the bottom panel shows the maximum multipole for the lensing spectrum.
    The largest multipole in the lensing spectrum is $\Lmax=999$, which is saturated at $\kmaxzero=0.3\hmpc$.
    }
    \label{fig:scale_cuts}
\end{figure}

As in cosmic shear, the CMB lensing spectrum is a line-of-sight projection of the matter power spectrum, so each lensing multipole receives contributions from a range of wavenumbers, and scale cuts can be derived by testing its sensitivity to modes with $k>\kmax$ \cite{derose2025lensingcounternarrativeeffective,Saraivanov:2026sxc}.
However, for the lensed temperature and polarization spectra, an additional convolution with the unlensed CMB spectra (see Eqs.~\ref{eq:delta_tt},~\ref{eq:delta_te}, and~\ref{eq:delta_ee}) mixes lensing multipoles, further complicating the relation between $\kmax$ and a multipole cut $\lmax$. Additionally, the lensing corrections $\Delta C^{XY}_{\ell}$ oscillate and/or cross zero at low $\ell$, making their fractional changes unsuitable for defining scale cuts. We therefore derive scale cuts using a different method than \cite{Saraivanov:2026sxc}, propagating variations of the matter power spectrum at $k>\kmax$ into the observables and assessing their impact using the likelihood covariance, as we now explain.

\begin{figure*}
    \centering
    \includegraphics[width=\textwidth]{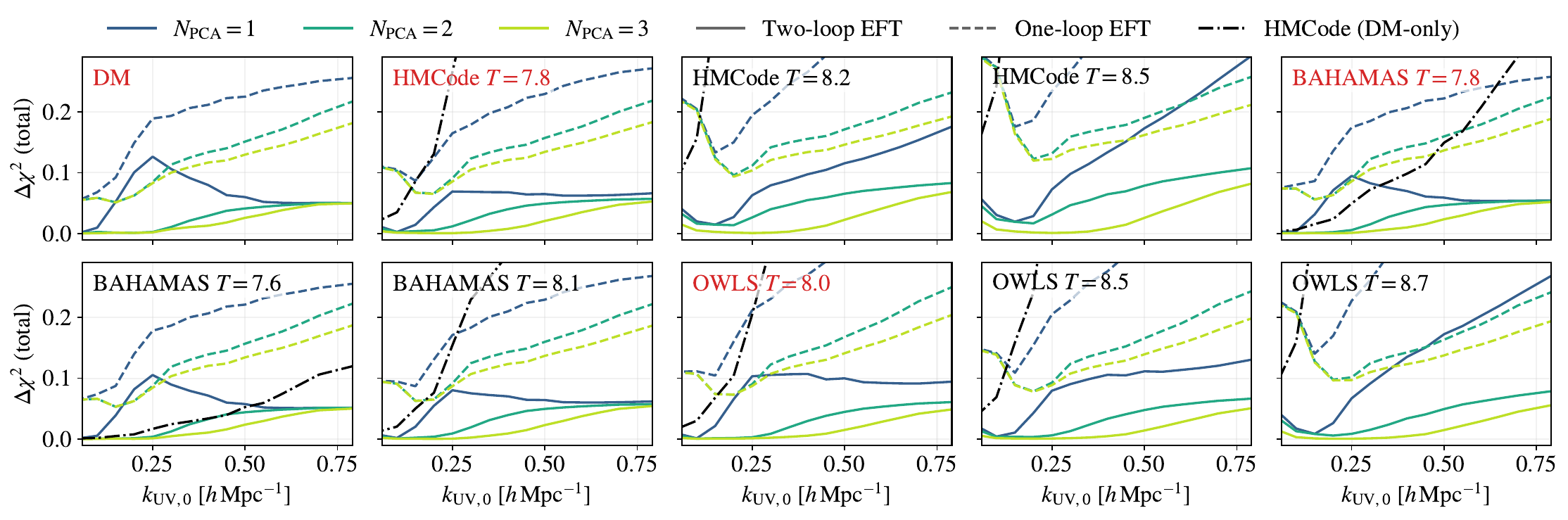}
    \caption{ The total (TTTEEE $+$ $\phi\phi$) $\Delta \chi^2$, measured against the fiducial data vector labelled inside the axes, as a function of $\kmaxzero$ for different analysis scenarios. The black dash-dotted line depicts the non-linear HMCode dark-matter-only model, whose $\Delta\chi^2$ vanishes against the dark-matter-only data vector by construction. The solid colored lines correspond to the two-loop EFTofLSS model, while the dashed colored lines correspond to the one-loop EFTofLSS. The color denotes the number of principal components that are allowed to vary. We use $T \equiv \log_{10}[T_{\rm AGN}/(1\,{\rm K})]$ as a shorthand notation. The models used for the scale cut calibration are indicated with a red label inside each panel.
    }
    \label{fig:chi2}
\end{figure*}

Our theory model, described in Sec.~\ref{sec:eftoflss}, requires a reference non-linear halo model to specify the power spectrum at UV wavenumbers through $R_{\rm ref.}$ in Eq.~\ref{eq:extrap}. Note that this also assumes Eq.~\ref{eq:kmaxscaling} for the scaling of $\kmax$ with the redshift.
This reference model provides a physically reasonable amplitude and shape for the power spectrum beyond the perturbative regime.
However, the choice of $R_{\rm ref.}$ is not unique, so we choose scale cuts that make the observables insensitive to plausible variations of this choice. This procedure can be interpreted as assigning a theoretical uncertainty to the power spectrum at $k>\kmax$ and discarding higher multipoles until the likelihood is insensitive to variations within this uncertainty. In deriving these cuts, we assume no modelling error for the EFTofLSS at $k\leq\kmax$. Also note that this way of setting the scale cuts does not rely on the particular EFTofLSS description adopted at $k\leq\kmax$, since it tests sensitivity to uncertainties at $k>\kmax$.

We use the $\Lambda$CDM non-linear ratio as $R_{\rm ref.}$, even when the forecast data vectors include $w_0w_a$CDM cosmologies and baryonic feedback. We estimate the uncertainty associated with this choice by considering alternative reference spectra with departures from $w_0=-1$ and $w_a=0$, and with baryonic feedback. To gauge the uncertainty from dark energy, we consider reference spectra with $w_0 \in [-1.3,-0.7]$ and $w_a\in[-0.3,0.3]$. For baryonic feedback, we consider three scenarios:
HMCode with $T = 7.8$, BAHAMAS with $T = 7.8$, and OWLS-AGN with $T = 8.0$, where we are using $T \equiv \log_{10}[T_{\rm AGN}/(1\,{\rm K})]$ as a shorthand notation. 
For each $\kmaxzero$, we determine scale cuts separately for the TTTEEE and lensing likelihoods, requiring the $\Delta\chi^2$ induced by each tested variation of the UV spectrum to be less than $0.2$ in each likelihood~\cite{DES:2021rex}. This procedure determines $\lmax$, the maximum multipole retained in the TTTEEE likelihood, and $\Lmax$, the highest multipole retained in the lensing likelihood, which are shown in the upper and lower panels of Fig.~\ref{fig:scale_cuts}, respectively. Note that the lensing cut saturates at $\Lmax=999$ when $\kmaxzero\geq0.3\hmpc$ once all available lensing bins are retained. 

We note two properties of our scale cut calibration.
First, the fiducial cosmology of our forecast, shown in Tab.~\ref{tab:fiducial_cosmology}, lies outside the $w_a$ range used to calibrate the scale cuts, but lies close to an approximate degeneracy in the $w_0$-$w_a$ plane along which the linear growth history remains close to that of $\Lambda$CDM, motivating this extrapolation.
Second, the OWLS-AGN models with $T = 8.5$ and $8.7$, also considered in Sec.~\ref{sec:forecast}, produce deviations that exceed our adopted theoretical error budget. These models therefore test the robustness of our scale cuts to feedback scenarios beyond those used in their calibration, as we discuss in Sec.~\ref{sec:forecast}.

\section{Simons Observatory forecast}
\label{sec:forecast}

At the sensitivity of Simons Observatory and future CMB experiments, baryonic feedback and inaccurate modelling of non-linear structure formation can bias several cosmological parameters by more than $0.3\sigma$ \cite{McCarthy_2022}. We investigate whether the one- and two-loop EFTofLSS can model mildly non-linear scales and mitigate these biases. After introducing the Fisher formalism, we assess the goodness of fit at fixed cosmology, relate the EFTofLSS principal components to baryonic feedback, and test for overfitting through parameter stability. Finally, we forecast cosmological constraints and biases after marginalizing over the EFTofLSS parameters, using the Simons Observatory multifrequency TT, TE, and EE likelihood together with the CMB lensing likelihood.

\subsection{Fisher formalism}

We use the Fisher formalism to forecast parameter constraints and estimate the biases induced by mismatches between the model and the data. The parameter displacement is defined as
\begin{equation}
    \boldsymbol{\delta} \equiv \Theta_{\rm MAP} - \Theta_{\rm fid}\,.
\end{equation}
Here, $\Theta$ contains all free cosmological and EFTofLSS parameters, $\Theta_{\rm MAP}$ denotes their maximum-a-posteriori values, and $\Theta_{\rm fid}$ denotes their fiducial values. For a Gaussian likelihood with data vector $\boldsymbol{d}$ and covariance $\Sigma$ held fixed, and a Gaussian prior, linearizing the model around $\Theta_{\rm fid}$ gives
\begin{equation}
    \boldsymbol{\delta} = F_{\mathcal{P}}^{-1}\left[ J^{T} \Sigma^{-1} \bigl(\boldsymbol{d} - \boldsymbol{m}_{\rm fid}\bigr) - F_{\Pi}(\Theta_{\rm fid} - \boldsymbol{\mu}) \right],
\end{equation}
where for the noiseless data vectors considered here, $\boldsymbol{d}-\boldsymbol{m}_{\rm fid}$ represents the model mismatch at the fiducial parameters.
We also define the model Jacobian and the fiducial model data vector as
\begin{equation}
    J \equiv \left. \frac{\partial \boldsymbol{m}(\Theta)}{\partial \Theta} \right|_{\Theta = \Theta_{\rm fid}}, \qquad \boldsymbol{m}_{\rm fid} \equiv \boldsymbol{m}(\Theta_{\rm fid})\,,
\end{equation}
where the matrix $F_{\Pi}$ and vector $\boldsymbol{\mu}$ are the prior Fisher matrix and mean of the Gaussian prior, respectively. The posterior Fisher matrix is
\begin{equation}
    F_{\mathcal{P}} = J^T \Sigma^{-1}J + F_{\Pi}\,,
\end{equation}
and the marginalized uncertainty on parameter $\Theta_i$ is
\begin{equation}
    \sigma_i = \sqrt{\bigl(F_{\mathcal{P}}^{-1}\bigr)_{ii}}\,.
\end{equation}

\subsection{Fit to the data}

Before performing a full parameter forecast, we focus here on the perturbative reach of the EFTofLSS applied to CMB lensing for fixed cosmology. For that, we perform a Fisher forecast for the counterterm spline coefficients alone, keeping the cosmological parameters fixed to their fiducial values in Tab.~\ref{tab:fiducial_cosmology}. In this section, we assess the goodness of fit through the corresponding $\Delta\chi^2$. In Sec.~\ref{sec:overfitting}, we also assess potential overfitting through the drift in the PCA amplitudes as a function of $\kmaxzero$.

In Fig.~\ref{fig:chi2}, we show the $\Delta\chi^2$ as a function of $\kmaxzero$ for each fiducial data vector, using different numbers of principal components and both one- and two-loop EFTofLSS modelling. The black dash-dotted curves show the non-linear HMCode prediction without baryonic feedback (DM-only). The curves flatten around $\kmaxzero=0.3\hmpc$, which is the point where the lensing information is fully saturated as shown in Fig.~\ref{fig:scale_cuts}. The increase in $\Delta\chi^2$ at very low values of $\kmaxzero$, particularly for the one-loop EFTofLSS, arises from the difference between the $\Lambda$CDM extrapolation and the true $w_0w_a$CDM model, which our scale cuts only guarantee to differ by $\Delta\chi^2<0.2$.

Considering the overall behaviour across the different feedback scenarios in Fig.~\ref{fig:chi2}, the one-loop EFTofLSS with a single principal component fits the data with $\Delta\chi^2<0.2$ for $\kmaxzero\lesssim0.2\hmpc$ and adding a second principal component extends this range to approximately $\kmaxzero=0.5\hmpc$.\footnote{We do not directly compare the values of $\kmaxzero$ reported here with the expected one-loop validity range for three-dimensional galaxy clustering, $k\lesssim0.2\hmpc$ at $z=0$. CMB lensing involves a line-of-sight integral, and, given the $\lmax(\kmaxzero)$ and $\Lmax(\kmaxzero)$ relations illustrated in Fig.~\ref{fig:scale_cuts} and discussed in Sec.~\ref{sec:scale_cuts}, the scale cuts are imposed on the angular multipoles. The extrapolation procedure further complicates a direct comparison~\cite{Braganca:2020nhv}.}
Restricting the comparison to feedback scenarios within the theoretical error budget used to calibrate the scale cuts (indicated with red labels in Fig.~\ref{fig:chi2}) extends this range to approximately $\kmaxzero=0.65\hmpc$ with two principal components.

The two-loop EFTofLSS fits the data with a single principal component with $\Delta\chi^2<0.2$ throughout the range considered for all feedback scenarios considered for the calibration of scale cuts. For the stronger feedback scenarios not used for the calibration (indicated with black labels in Fig.~\ref{fig:chi2}), the same criterion holds for $\kmaxzero\lesssim0.6\hmpc$. Including two principal components allows all baryonic models to be fit with $\Delta\chi^2<0.2$ up to $\kmaxzero=0.8\hmpc$, the highest value we consider.

Overall, the one-loop EFTofLSS provides a noticeably better fit to the data than the dark-matter-only HMCode prediction, while the two-loop EFTofLSS improves the fit further. The two-loop model also achieves a comparable fit with fewer principal components, suggesting that the extra counterterms are needed for fitting the power spectrum on smaller scales. Since a single principal component provides a good fit for both the one- and two-loop models, we focus on results with one principal component for the remainder of this manuscript. Using one nuisance parameter also reduces the risk of overfitting the data, which we discuss in Sec.~\ref{sec:overfitting}.

Finally, we examine the physical interpretation of the PC directions by comparing them with the baryonic feedback parameters.
In HMCode, the impact of baryonic feedback on the matter power spectrum can be described by a single parameter, the effective AGN heating temperature, expressed as $T \equiv \log_{10}[T_{\rm AGN}/(1\,{\rm K})]$~\cite{Mead:2020vgs,Kilo-DegreeSurvey:2023gfr}.
To assess whether the EFTofLSS counterterm PCs are related to the AGN temperature, we compute the cosine similarity between the changes in the data vector induced by varying $T$ in HMCode and the amplitude of each EFTofLSS PC:
\begin{equation}\label{eq:cos_similarity}
    \cos\theta(T,\,PC) = \frac{
        [\partial_T \bar{m}_{\rm HMCode}] \cdot
        [\partial_{\rm PC} \bar{m}_{\rm EFT}]
    }{
        |\partial_T \bar{m}_{\rm HMCode}|
        |\partial_{\rm PC} \bar{m}_{\rm EFT}|
    }\,,
\end{equation}
where $\bar{m}(\Theta)=\Sigma^{-1/2}m(\Theta)$ denotes the normalized data vector, computed using either HMCode or EFTofLSS and the same fixed covariance $\Sigma$. We evaluate the HMCode derivative at $T=7.8$ and the EFTofLSS derivatives with all PC amplitudes set to zero, keeping the remaining parameters at their fiducial values.

\begin{figure}
    \centering
    \includegraphics[width=0.99\columnwidth]{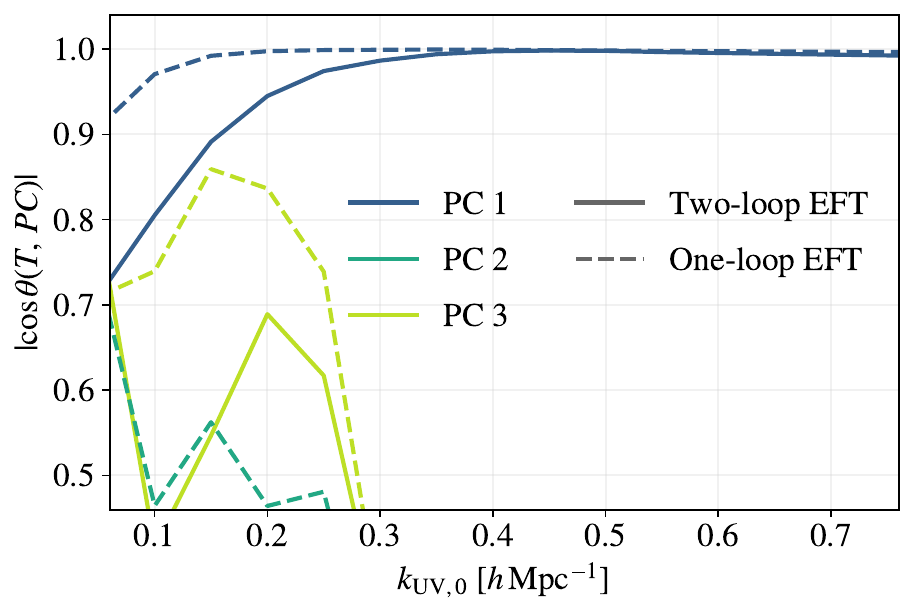}
    \caption{
    The cosine similarity, as defined in Eq.~\ref{eq:cos_similarity}, between the HMCode baryonic feedback model and the EFTofLSS principal components. The solid lines show the two-loop principal components and the dashed lines the one-loop, with the color denoting principal component number.
    }
    \label{fig:cos_similarity}
\end{figure}

The cosine similarity is shown in Fig.~\ref{fig:cos_similarity}. For both the one-loop and two-loop EFTofLSS, the first principal component shows the strongest alignment with the response to HMCode's baryonic feedback parameter. At lower $\kmaxzero$, the observables are less sensitive to baryonic feedback, and the counterterms instead capture other non-linear corrections to the dark-matter spectrum. The alignment with the baryonic response at higher $\kmaxzero$ can explain why a single EFTofLSS principal component is sufficient to mitigate baryonic feedback and provide a good fit to the data.

\subsection{Overfitting} \label{sec:overfitting}

Assessing parameter stability provides a diagnostic of possible overfitting~\cite{Foreman:2015lca}.
To quantify this stability, we examine how the fitted PCA amplitudes shift as we increase $\kmaxzero$ and include additional angular modes. We therefore compare a fit to the modes retained at a lower $\kmaxzero$ (dataset $1$) with a fit that also includes the additional modes using a higher $\kmaxzero$ (dataset $2$). We assess whether the resulting shift is larger than expected from statistical fluctuations using the \emph{parameter difference in update form} statistic, $Q_{\rm UDM}$~\cite{raveri_concordance_2019}, which accounts for the correlations between the two fits due to their shared data.
This statistic is defined as
\begin{equation}\label{eq:qudm}
    Q_{\rm UDM} = (\boldsymbol{\alpha}_{1} - \boldsymbol{\alpha}_{2})^{T} (\mathcal{C}_{1} - \mathcal{C}_{2})^{-1} (\boldsymbol{\alpha}_{1} - \boldsymbol{\alpha}_{2})\,,
\end{equation}
where $\boldsymbol{\alpha}_{1}$ and $\mathcal{C}_{1}$ are the posterior mean PCA amplitudes and their covariance at the smaller $\kmaxzero$, while $\boldsymbol{\alpha}_{2}$ and $\mathcal{C}_{2}$ are their updated values at the larger $\kmaxzero$. Since this statistic follows a $\chi^2$ distribution, with $d={\rm rank}(\mathcal{C}_{1}-\mathcal{C}_{2})$ degrees of freedom, the probability of obtaining a smaller value of the statistic is
\begin{equation}
p = \int\limits_0^{Q_{\rm UDM}} \chi^2(x, d)\,{\rm d}x\,,
\end{equation}
where $\chi^2(x,d)$ denotes the $\chi^2$ probability density. We express this cumulative probability as a Gaussian significance using
\begin{equation}\label{eq:nsigma}
    N_{\sigma} = \sqrt{2}\,\operatorname{erf}^{-1}(p)\,,
\end{equation}
and we adopt $N_{\sigma}>2$ as our threshold for significant parameter drift~\cite{Foreman:2015lca}.

\begin{figure}
    \centering
    \includegraphics[width=\columnwidth]{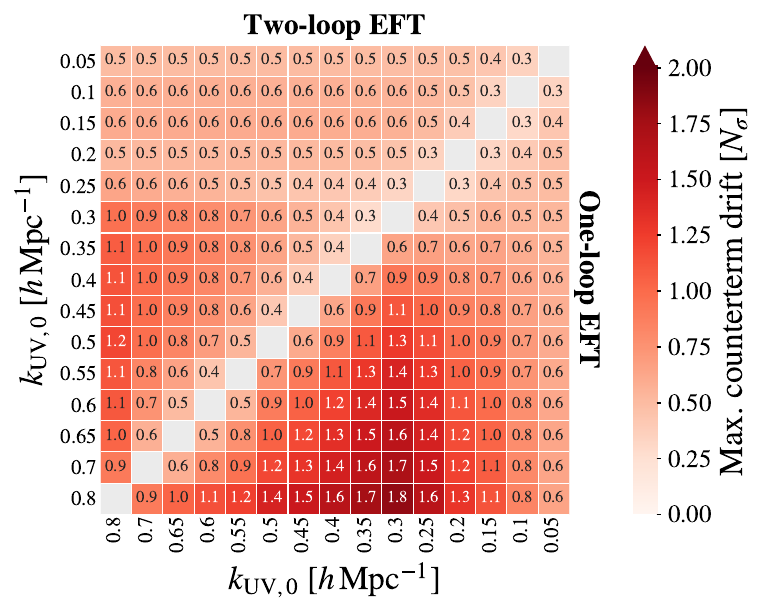}
    \caption{
    The drift in the principal component amplitude, in units of $N_{\sigma}$ as defined in Eq.~\ref{eq:nsigma}, as a function of $\kmaxzero$. The upper triangle contains the two-loop drift, and the lower triangle contains the one-loop drift. The axes show the $\kmaxzero$ measurements that are compared, where the larger value should always be interpreted as dataset $2$ in Eq.~\ref{eq:qudm}. The displayed $N_\sigma$ (see Eq.~\ref{eq:nsigma}) is the maximum shift observed over the dark-matter-measurement and {\it all} baryonic feedback measurements.
    }
    \label{fig:pca_drift}
\end{figure}

In Fig.~\ref{fig:pca_drift}, we show the maximum observed $N_{\sigma}$ over {\it all} of the baryonic feedback models for each pair of $\kmaxzero$ values. We stress this is a very conservative estimate, taking the worst-case scenario for each pair of $\kmaxzero$. For each pair, the larger $\kmaxzero$ corresponds to the updated dataset, labelled $2$ in Eq.~\ref{eq:qudm}. In no case do we see any overfitting at greater than $2\sigma$. For the two-loop EFTofLSS, the drift remains below $1\sigma$ for $\kmaxzero \lesssim 0.65\hmpc$, while for the one-loop EFTofLSS, this range reduces to $\kmaxzero \lesssim 0.4\hmpc$. 
Note that the parameter shifts are not monotonic as $\kmaxzero$ increases. This is for two reasons. The first is that the scale dependence is quite moderate, meaning the shift between adjacent $\kmaxzero$ is small compared to the error bars. The second is that the error bars at the lower $\kmaxzero$ are significantly larger than the error bars at larger $\kmaxzero$, making the lower values statistically compatible with all measurements.

Therefore, we find no evidence for overfitting in our results. This reflects the limited additional information available at high multipoles for CMB lensing, since at larger multipoles, SO instrumental noise and non-lensing secondary CMB signals contribute substantially. Hence, extending the analysis to smaller scales does not necessarily yield substantially tighter constraints.
Moreover, we checked whether some of the observed drift in Fig.~\ref{fig:pca_drift} arises from the $\kmaxzero$-dependent extrapolation scheme, since the separation of scales is not exact for projected fields. The counterterms may absorb contributions to the observables that change with $\kmaxzero$, potentially inducing scale dependence in the fitted amplitudes. 
We isolate this effect by repeating the test with the extrapolation scheme fixed by the larger $\kmaxzero$ for both fits, while using the smaller $\kmaxzero$ to determine the $\lmax$ and $\Lmax$ cuts for the reduced dataset. We find that fixing the extrapolation scheme reduces the observed shifts by up to a factor of $5$, bringing all shifts below $0.5\sigma$, indicating even less evidence for overfitting.

\subsection{Parameter forecasts}

\begin{figure*}
    \centering
    \includegraphics[width=\textwidth]{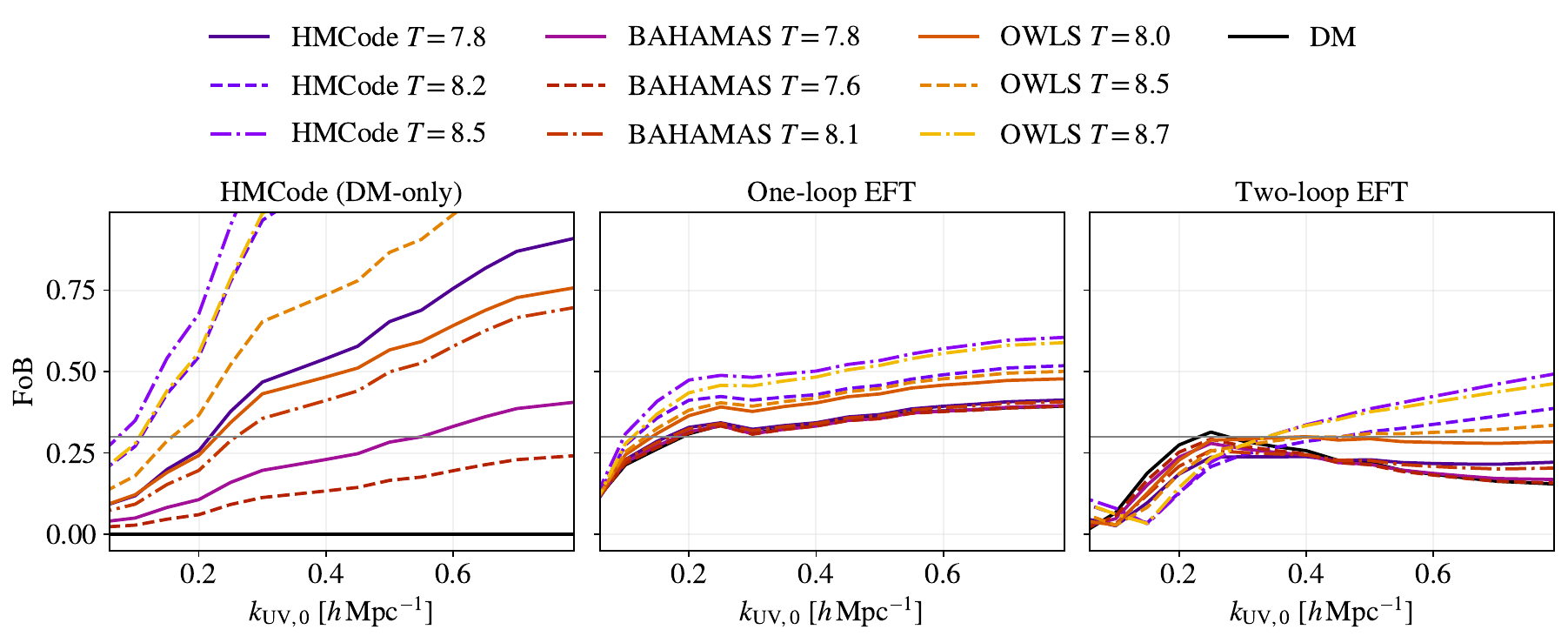}
    \caption{
    The figure of bias, defined in Eq.~\ref{eq:fob}, as a function of $\kmaxzero$ for each of the data vectors and each of the non-linear modelling choices. The left panel shows the HMCode non-linear model, the center panel shows the one-loop EFTofLSS, and the right panel shows the two-loop EFTofLSS. The dark gray line shows the $0.3\sigma$ threshold. The solid lines are the models used for scale cuts in Sec.~\ref{sec:scale_cuts}. 
    }
    \label{fig:fob}
\end{figure*}

Having addressed the potential for overfitting, we now vary the cosmological parameters and forecast their constraints and biases using the Fisher formalism, comparing the dark-matter-only HMCode model with our one-loop and two-loop EFTofLSS models.
We summarize the results in the full eight-dimensional cosmological parameter space using the figure of merit and figure of bias. A parameter-by-parameter breakdown of these results can be found in App.~\ref{sec:parameter_results}.
The figure of merit is defined as 
\begin{equation}\label{eq:fom}
    {\rm FoM} = 1/\sqrt{{\rm det}(\mathcal{C}_{\mathcal{P}(\Theta)})} \,,
\end{equation}
where $\mathcal{C}_{\mathcal{P}(\Theta)}$ is the posterior covariance matrix of the cosmological parameters, after marginalizing over the PC amplitudes for the EFTofLSS models. The figure of bias is defined as
\begin{equation}\label{eq:fob}
    \text{$\displaystyle {\rm FoB} = \sqrt{(\Theta_{\rm MAP} - \Theta_{\rm fid})^{T}[\mathcal{C}_{\mathcal{P}(\Theta)}]^{-1}(\Theta_{\rm MAP} - \Theta_{\rm fid})} $}\,.
\end{equation}

In Fig.~\ref{fig:fob}, we show the FoB of the HMCode non-linear model and the one-loop and two-loop EFTofLSS models, each EFTofLSS model using one principal component. We find a strong variation in the FoB for HMCode across the different feedback scenarios. For strong feedback models, the dark-matter-only HMCode model reaches the adopted ${\rm FoB}=0.3$ threshold already at $\kmaxzero\simeq0.05\hmpc$. The one-loop EFTofLSS modestly extends the usable range to $\kmaxzero\simeq0.1\hmpc$. For the feedback models included in the scale-cut calibration in Sec.~\ref{sec:scale_cuts}, denoted by solid lines in Fig.~\ref{fig:fob}, however, both HMCode and the one-loop EFTofLSS remain below this threshold up to $\kmaxzero\simeq0.2\hmpc$. We also find that the different curves are closer for the one-loop EFTofLSS result, indicating that part of the baryonic feedback is being absorbed by the counterterm in Eq.~\ref{eq:1lct}. The two-loop EFTofLSS performs substantially better, with its FoB remaining below $0.3$ for $\kmaxzero\lesssim0.25\hmpc$ across all baryonic scenarios and staying close to this threshold up to $\kmaxzero\simeq0.35\hmpc$. For the feedback models included in the scale-cut calibration, the two-loop EFTofLSS remains below the threshold up to $\kmaxzero=0.8\hmpc$.

Examining the parameter-by-parameter biases (see App.~\ref{sec:parameter_results}), we find the largest HMCode biases in $\log(10^{10}A_s)$, $\Omega_{\rm c}h^2$, and $\tau$, with shifts of up to $0.4\sigma_{\rm ref}$. These biases are reduced to $\lesssim 0.2\sigma_{\rm ref}$ when using the EFTofLSS, regardless of the loop order.
Additionally, $\Omega_{\rm b}h^2$ and $w_a$ show moderate biases that almost vanish when using one- and two-loop EFTofLSS modelling.

We caution, however, against interpreting these results as establishing the validity of the two-loop EFTofLSS up to $\kmaxzero=0.8\hmpc$ at redshift $z=0$, since this apparent reach depends on the relation mapping $\kmaxzero$ to the angular scale cuts $\lmax$ and $\Lmax$, as described in Sec.~\ref{sec:scale_cuts}. Moreover, since the CMB lensing information largely saturates for $\kmaxzero>0.3\hmpc$ (see Fig.~\ref{fig:scale_cuts}), increasing $\kmaxzero$ adds little constraining power, as we discuss below, limiting the effects of the additional small-scale modes on the fit. Cross-correlations with other datasets and higher-order statistics could provide further tests of perturbative validity by probing scale dependence that is less accessible in the smooth projected two-point function. We plan to investigate these extensions in future work.

\begin{figure}
    \centering
    \includegraphics[width=\columnwidth]{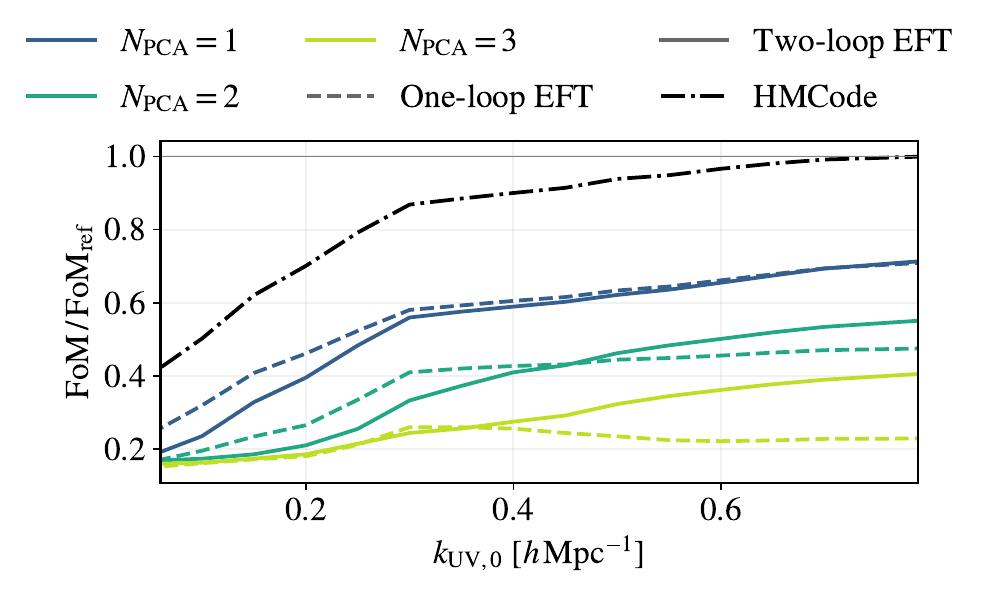}
    \caption{
    The figure of merit (FoM) as defined in Eq.~\ref{eq:fom} of the eight cosmological parameters and normalized to the HMCode FoM at $\kmaxzero=0.8\hmpc$ as a function of $\kmaxzero$ for each of the non-linear models. For the one-loop and two-loop EFT, we show the figure of merit for up to three principal components. 
    }
    \label{fig:fom}
\end{figure}

In Fig.~\ref{fig:fom}, we show the FoM of the HMCode non-linear model and the one- and two-loop EFTofLSS models, using one, two, and three principal components for the EFTofLSS models. The FoM is normalized to that of HMCode at $\kmaxzero=0.8\hmpc$.  The sharp feature seen at $\kmaxzero \simeq 0.3\hmpc$ is a result of the lensing data being saturated at that scale, as seen in Fig.~\ref{fig:scale_cuts}, limiting the FoM improvement from that scale onward. At fixed $\kmaxzero$, both EFTofLSS models lose about $20$--$40\%$ in the FoM relative to HMCode through marginalization over the counterterm amplitudes.
However, as noted above, the dark-matter-only and the EFTofLSS models become biased at different scales.
The dark-matter-only HMCode model reaches ${\rm FoB}=0.3$ at $\kmaxzero\simeq0.05\hmpc$ (or $\kmaxzero\simeq0.2\hmpc$ for the feedback models included in the scale-cut calibration), where its relative FoM is about $40\%$ ($70\%$). At the largest scale cuts satisfying the same bias criterion, the one-loop EFTofLSS reaches a relative FoM of roughly $32\%$ ($46\%$), below the corresponding HMCode values. The two-loop EFTofLSS instead reaches a relative FoM of roughly $56\%$ ($72\%$) under the same criterion. We further notice the EFTofLSS results are more robust across the different feedback models. The two-loop EFTofLSS's access to smaller scales more than compensates for the loss in constraining power from marginalizing over the counterterm, yielding stronger joint constraints on the eight cosmological parameters than HMCode.

Examining the marginalized uncertainties of individual parameters (see App.~\ref{sec:parameter_results}) shows that, when using one PC, the largest losses in constraining power occur for $\Omega_{\rm b}h^2$, $\log(10^{10}A_s)$, $\tau$, and $\Omega_{\rm c}h^2$, although these losses remain small. These trends are consistent with the known effects of marginalizing over baryonic feedback~\cite{McCarthy_2022}. The constraints on $w_0$, $w_a$, and $n_s$ remain comparable to those from HMCode for $\kmaxzero\gtrsim0.25\hmpc$ and the constraining power on $H_0$ is largely unaffected by the EFTofLSS, as expected given the tight constraints provided by the combination of the CMB acoustic peak positions and the BAO and supernova priors.

\section{Conclusion}

Future CMB lensing measurements will become increasingly sensitive to non-linear structure formation, baryonic physics, and possible signatures of new physics. Hence, consistent theoretical modelling is necessary for robust cosmological parameter inference. In this manuscript, we outline a procedure for applying the EFTofLSS to CMB observables and forecast cosmological parameter constraints for the Simons Observatory. 

The lensing spectrum involves an integral over $k$ and therefore receives contributions from $k>k_{\rm NL}$ even on mildly non-linear angular scales, requiring a prescription for the power spectrum beyond the regime of validity of the EFT. We extrapolate the power spectrum using the corresponding $\Lambda$CDM non-linear ratio and apply scale cuts calibrated by varying $w_0$, $w_a$, and baryonic feedback to limit sensitivity to uncertain UV contributions. This framework closely follows the one used for galaxy weak lensing~\cite{Saraivanov:2026sxc,Chen:2026usz}.

Modelling the time dependence of the EFTofLSS counterterms naively introduces $\mathcal{O}(10)$ free parameters per counterterm.
By performing a PCA, we find that only a few linear combinations of these parameters can be constrained by the data. We demonstrate that, at either one- or two-loop order, varying only the amplitude of the leading principal component is sufficient to reduce the biases in the marginalized one-dimensional cosmological parameter constraints to below about $0.2\sigma$, with only a few percent loss of constraining power per parameter relative to HMCode at fixed $\kmaxzero$. We find this single parameter to be strongly aligned with $T_{\rm AGN}$.

The one- and two-loop EFTofLSS yield similar marginalized cosmological parameter constraints at a fixed $\kmaxzero$, but the one-loop EFTofLSS provides a good fit to the data (that is, $\Delta\chi^2<0.2$) up to $\kmaxzero\approx0.2\hmpc$, while the two-loop EFTofLSS provides a good fit up to $\kmaxzero\approx0.6\hmpc$.
At a fixed scale, marginalizing over this parameter results in a $\sim30\%$ drop in the figure of merit compared to the dark-matter-only HMCode model.
However, on scales where each model is unbiased in the full parameter space, the two-loop EFTofLSS is able to exceed the figure of merit of the dark-matter-only model by a few percent for weaker baryonic feedback models and by $40\%$ for stronger feedback models.
The one-loop EFTofLSS still loses about $20\%$ in the figure of merit on scales where it is unbiased.
The two-loop model also appears to absorb the effects of different baryonic feedback prescriptions more smoothly as smaller scales are included. 

Future CMB lensing observations with lower noise and cross-correlations with other large-scale structure observables~\cite{Farren:2023yna,ACT:2024npz,ACT:2023skz} will place increasingly stringent demands on theoretical modelling. Extending the range of scales accessible to the EFTofLSS could also open new opportunities to probe physics beyond $\Lambda$CDM, including larger ultralight axion masses~\cite{Lague:2026sbd,Verdiani:2026cpc}. By allowing counterterm amplitudes to be inferred directly from the data, the EFTofLSS offers a systematic framework for exploring these cosmologies. Our work thus paves the way for applying the one- and two-loop EFTofLSS to CMB lensing and for consistently modelling its cross-correlations with galaxy lensing and other large-scale structure observables.

\section*{Acknowledgements}

The authors would like to thank Stony Brook Research Computing and Cyberinfrastructure and the Institute for Advanced Computational Science at Stony Brook University for access to the high-performance SeaWulf computing system, which was made possible by $\$1.85$M in grants from the National Science Foundation (awards 1531492 and 2215987) and matching funds from the Empire State Development’s Division of Science, Technology and Innovation (NYSTAR) program (contract C210148). VM is supported by the Roman Project Infrastructure Team ``Maximizing Cosmological Science with the Roman High Latitude Imaging Survey''  (NASA contracts 80NM0018D0004-80NM0024F0012). VM is also partially supported by the Roman Project Infrastructure Team ``A Roman Project Infrastructure Team to Support Cosmological Measurements with Type Ia Supernovae'' (NASA contract 80NSSC24M0023).

\appendix

\bibliographystyle{apsrev4-2}
\bibliography{bib}

\section{CAMB settings} \label{app:CAMBsettings}

Analyses for Simons Observatory and other CMB observatories require higher numerical accuracy in the Boltzmann solver.
We use the CAMB settings from~\cite{zhu2025attentionbasedneuralnetworkemulators} (originally based on~\cite{MacInnis:2023vif}), which were selected to achieve accuracy at the level of cosmic variance.
We summarize these settings in Tab.~\ref{tab:camb_settings}.
\begin{table} [h]
    \centering
    \renewcommand{\arraystretch}{1.3}
    \begin{tabular}{lr}
        \hline
        CAMB setting & Value \\\hline
        Accuracy Boost & 1.5 \\
        $\ell$ Accuracy Boost & 3.0 \\
        $\ell$ Sample Boost & 1.5 \\
        $\ell_{\rm max}$ & 15000 \\
        Lens Potential Accuracy & 30 \\
        \hline
    \end{tabular}
    \caption{The CAMB accuracy settings used for the forecast in Sec.~\ref{sec:forecast}. Note that $\ell_{\rm max}$ can numerically alter the CMB spectra at other multipoles.}
    \label{tab:camb_settings}
\end{table}

\section{PCA details}

\begin{figure}[h]
    \centering
    \includegraphics[width=0.99\columnwidth]{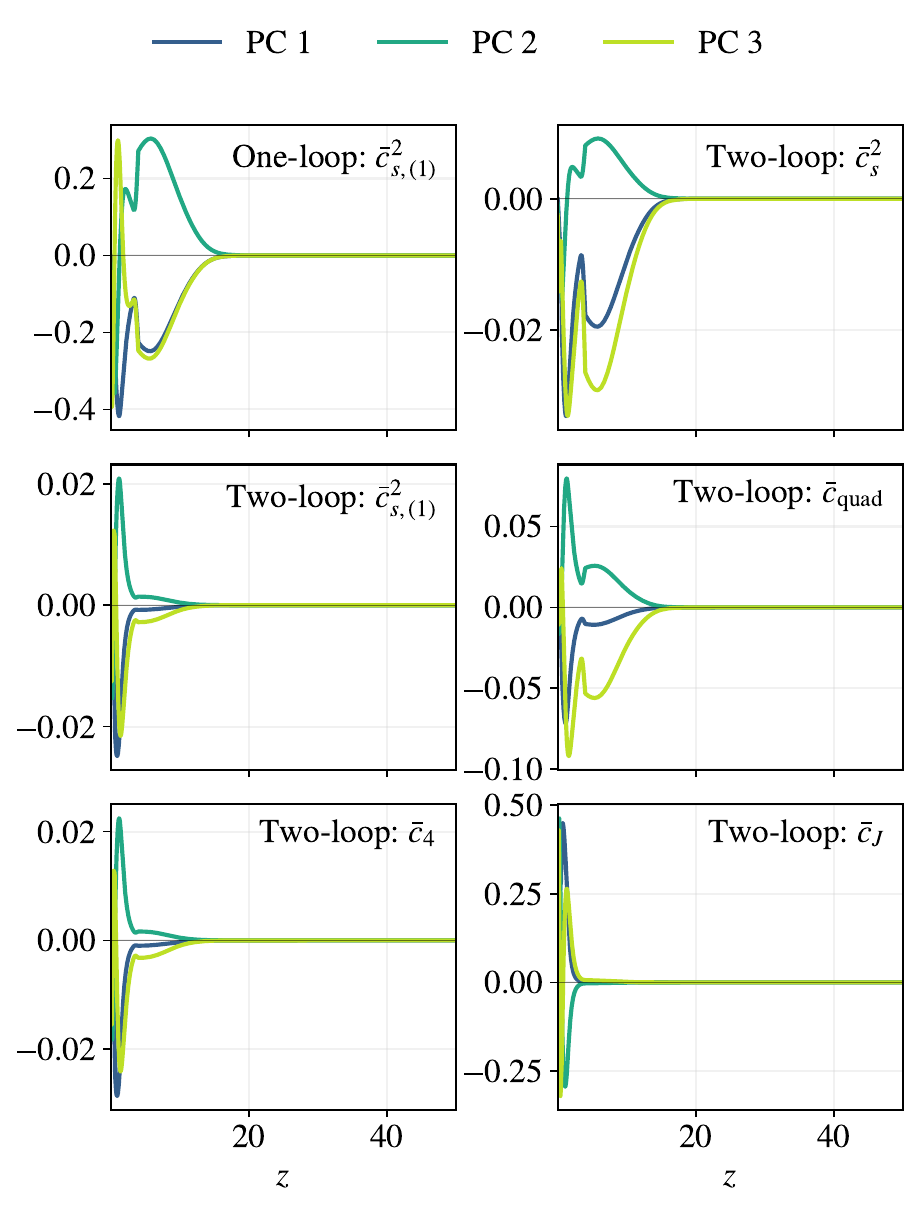}
    \caption{
    The first three principal components projected onto the counterterms in Eq.~\ref{eq:counterterms}.
    }
    \label{fig:pca}
\end{figure}

In Fig.~\ref{fig:pca}, we show the first three principal components projected onto the counterterms of Eq.~\ref{eq:counterterms} as a function of redshift. For $z\gtrsim20$, these components become negligible. The principal components have their largest amplitudes at low redshift, $z<5$, which is consistent with the redshift range that provides most of the support of the CMB lensing kernel.

\section{EFTofLSS Emulators}\label{sec:emulator}

We use neural networks to emulate each of the terms needed for the two-loop EFTofLSS expressions. The approach here differs from the one in~\cite{Saraivanov:2026sxc} because CAMB requires the non-linear ratio to be supplied rather than the full power spectrum. This paper also requires the model to account for $w_0w_a$CDM. 

We construct a model that combines neural networks and polynomial chaos expansion~\cite{Reboucas:2025hue}, which learns the mapping
\begin{equation}
    (A_s\times10^{9}, n_s, H_0, \Omega_{\rm b}, \Omega_{\rm m}, w_0, w_a) \mapsto \sinh^{-1} \left( \frac{X(k)}{N(k)} \right)\,.
\end{equation}
Here, $X(k)$ denotes the term to be emulated, and $N(k)$ is a normalization spectrum, which we choose to be different depending on $X(k)$. We adopt the IR-resummation scheme from~\cite{Garny_2022}, defining $P_{0}^{mL}$ as the damped linear power spectrum entering at $m$-th loop order, and $P_{n}^{mL}$ as the $n$-th loop correction computed using $P_{0}^{mL}$. Thus we need three damped linear spectra, two damped one-loop spectra, a single damped two-loop spectrum, and a damped quadratic counterterm spectrum. Since CAMB requires the non-linear \emph{ratio}, we must also learn $k^4/P_{\rm L}(k)$, the ratio of the stochastic term to the linear spectrum output by CAMB. 

\begin{figure*}
    \centering
    \includegraphics[width=\textwidth]{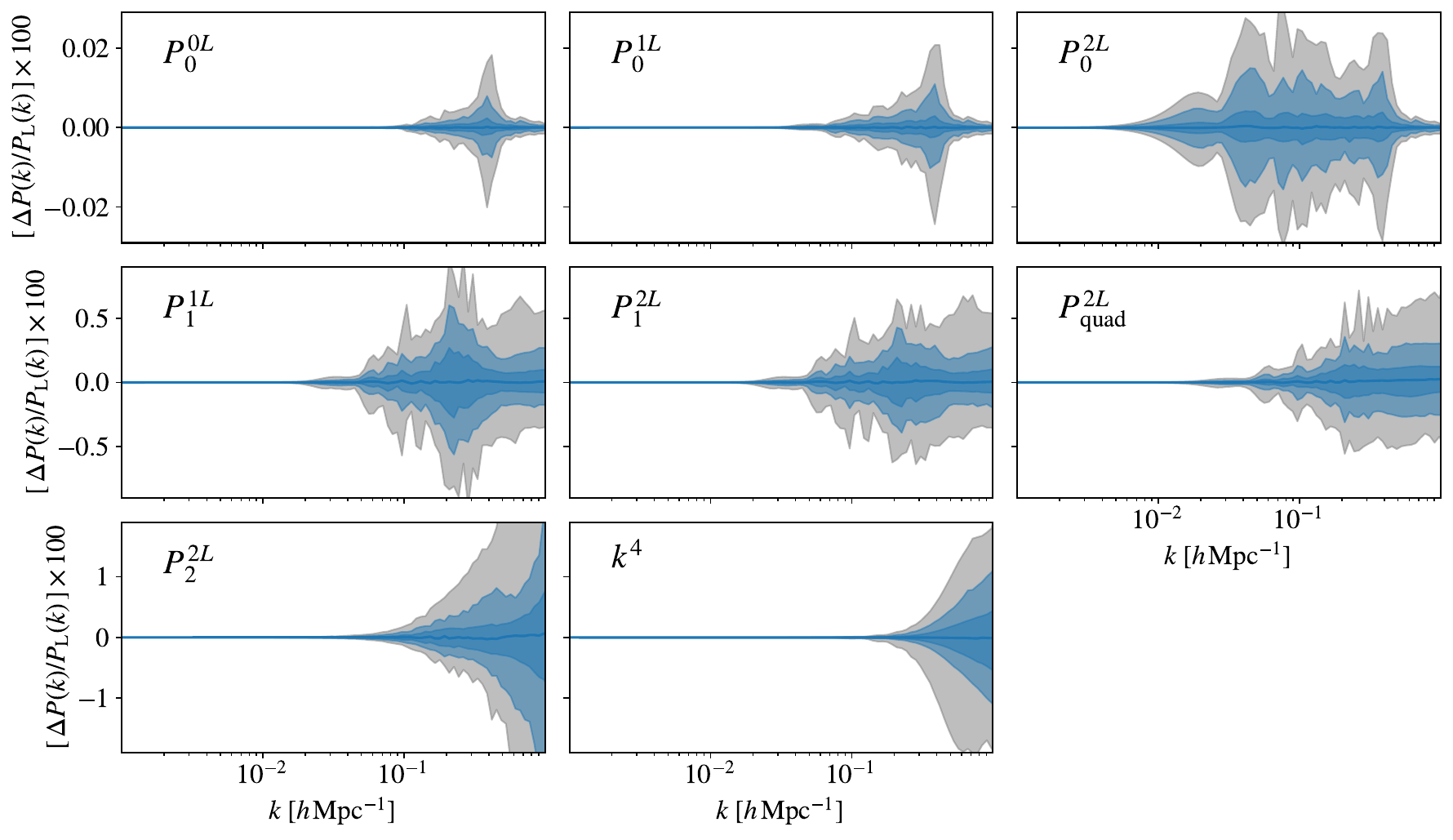}
    \caption{
    The emulation errors as a function of $k$ for each of the terms needed for the two-loop EFTofLSS non-linear ratio. The shaded areas show the $68\%$, $95\%$, and $99.7\%$ intervals across the testing set.
    }
    \label{fig:emulation_errors}
\end{figure*}

For the normalization spectra $N(k)$, we use a different choice depending on the term to emulate. For the linear, one-loop, quadratic, and stochastic terms, we use $N(k)=P_{\rm L}(k)$, with $P_{\rm L}(k)$ as output by CAMB (i.e.,~no damping). For the two-loop term, we use $N(k)=P_{\rm L}+P_{1}^{1L}$. These were chosen empirically to give the highest emulation accuracy. The two-loop term must also be scaled by $1+(P_1^{1L}/P_{\rm L})$ before being passed to CAMB. We tested that using the emulated $P_1^{1L}$ to un-normalize the emulated two-loop term has a negligible effect on the emulation errors. In Fig.~\ref{fig:emulation_errors}, we show the difference between the emulated $X(k)$ and the true $X(k)$, normalized by the linear spectrum as output by CAMB. We find that the emulation errors increase with loop order, but remain smaller than $2\%$ in the range of $k$ we consider in this manuscript.

\section{Parameter breakdown}
\label{sec:parameter_results}

\begin{figure*}
    \centering
    \includegraphics[width=\textwidth]{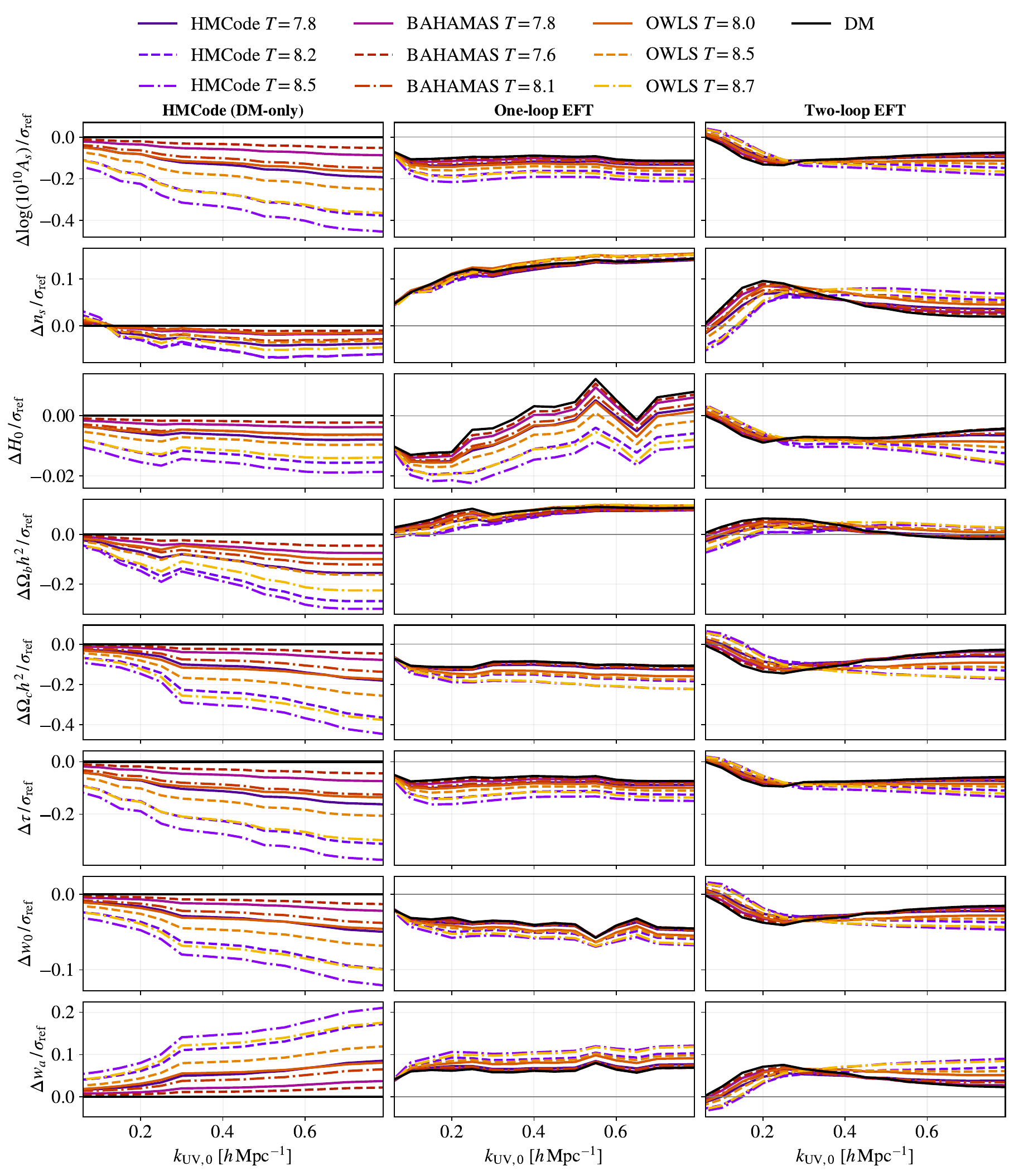}
    \caption{
    The one-dimensional parameter shifts, in units of $\sigma_{\rm ref}$ from Tab.~\ref{tab:fiducial_cosmology}, of each of the eight cosmological parameters as a function of $\kmaxzero$. 
    The first column shows the parameter shifts for the HMCode dark-matter-only model. The middle column shows the shifts for the one-loop EFTofLSS with one principal component and the right column shows the shifts for the two-loop EFTofLSS with one principal component. The line colors correspond to different baryonic feedback scenarios. The solid lines denote the models used for scale cuts in Sec.~\ref{sec:scale_cuts}.
    }
    \label{fig:hm_shifts}
\end{figure*}

\begin{figure*}
    \centering
    \includegraphics[width=\textwidth]{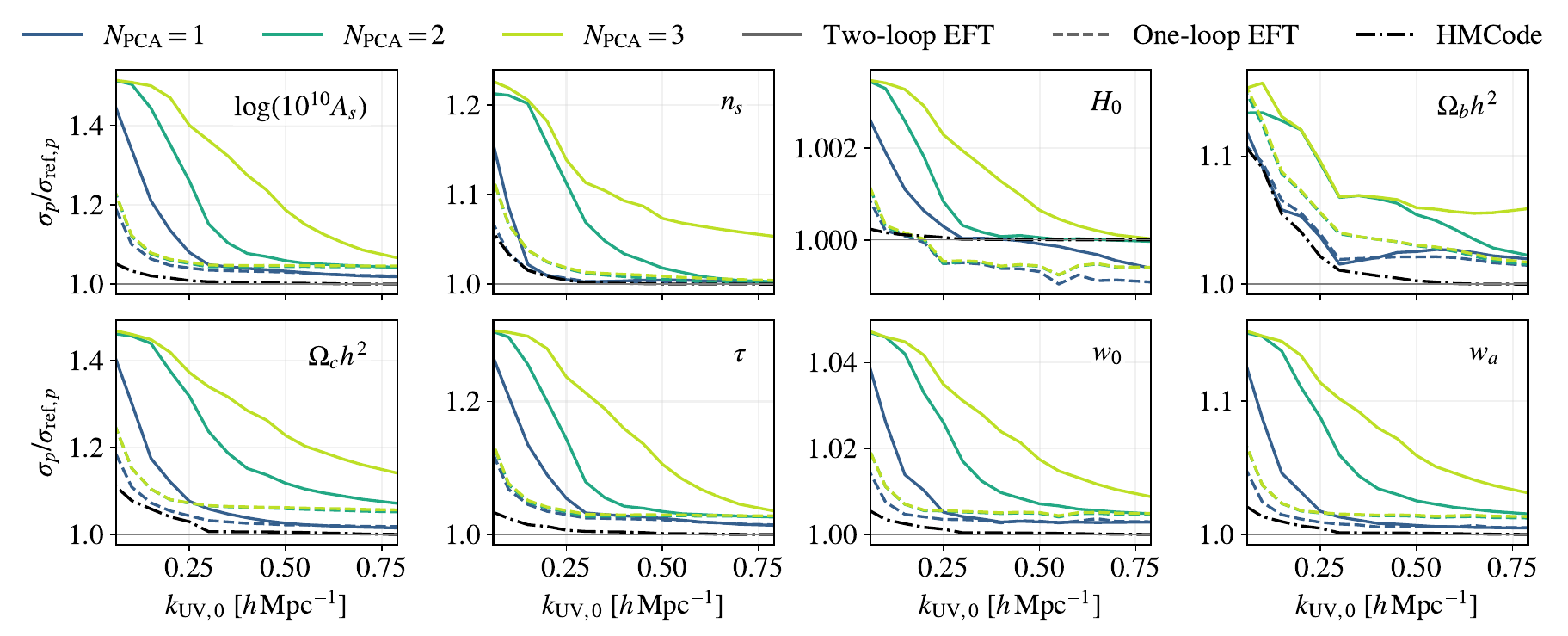}
    \caption{
    The marginalized constraints on each of the eight cosmological parameters as a function of $\kmaxzero$. The dash-dotted line is for the HMCode dark-matter-only model, the dashed lines are for the one-loop EFTofLSS, and the solid lines are for the two-loop EFTofLSS. For the EFTofLSS models, the colors denote the number of principal components. 
    }
    \label{fig:errorbars}
\end{figure*}

In this appendix, we show the biases and marginalized uncertainties of individual cosmological parameters.
Fig.~\ref{fig:hm_shifts} shows the shift in each of the eight cosmological parameters in units of the reference uncertainties given in Tab.~\ref{tab:fiducial_cosmology}. The middle and last columns show the results for the one-loop and two-loop EFTofLSS, respectively.
Each row corresponds to a different cosmological parameter, with all remaining parameters marginalized over.
In Fig.~\ref{fig:errorbars}, we show the marginalized uncertainty for each of the eight cosmological parameters.

\end{document}